# Growth and Raman spectroscopy of thickness-controlled rotationally faulted multilayer graphene

H. Kato, N. Itagaki, H. J. Im[*]

*Graduate School of Science and Technology, Hirosaki University, Hirosaki 036-8561, Japan*

**Abstract**

We report the growth of thickness-controlled rotationally faulted multilayer graphene (rf-MLG) on Ni foils by low-pressure chemical vapor deposition and their characterization by micro-Raman spectroscopy. The surface morphology and thickness were investigated by scanning electron microscope, X-ray diffraction, and transmittance measurements. These results have revealed that the thickness of rf-MLG can be effectively controlled by the thickness of the Ni foil rather than the flow rate of $CH_4$, $H_2$, Ar. In the Raman spectroscopy measurements, we observed most Raman peaks of the graphitic materials. Raman spectra can be categorized into four patterns and show systematic behaviors. Especially, the in-plane (~1880 $cm^{-1}$, ~2035 $cm^{-1}$) and out-of-plane (~1750 $cm^{-1}$) modes are successfully analyzed to explain the dimensionality of rf-MLG as in the twisted (or rotated) bilayer graphene. In addition, it is found that the two peaks at ~1230 $cm^{-1}$ and ~2220 $cm^{-1}$ well reflect the properties of the in-plane mode. The peak intensities of the above four in-plane modes are proportional to that of 2D band, indicating that they share the common Raman resonance process.

[*]Corresponding author
Email address: hojun@hirosaki-u.ac.jp (H. J. Im)

# 1. Introduction

Graphene has been much attracted in both fundamental science and application due to outstanding electrical and optical properties, which come from the typical electronic band structure, the so-called Dirac-cone [1, 2, 3]. In graphene, electrons behave like massless particles with charge and spin owing to the linearly dispersive electric band near the Fermi level [4]. Particularly, its high carrier mobility and absorption coefficient have been expected to make graphene the substitution for silicon in the opto-electronic devices [5]. However, to this end, there are some problems we have to resolve. When two single-layer graphenes (SLGs) are stacked and form bilayer graphene (BLG), the above outstanding properties are reduced by the deformation of the Dirac-cone. As the layer number increases, the three-dimensional properties become stronger, and graphene eventually changes into graphite. In general, it has been accepted that the Bernal (AB-) stacking more than ~10 graphene layers gives rise to the graphitization [6]. Fortunately, it has been recently reported that the rotational stacking can dramatically improve the physical properties of graphene [7, 8]. These have been intensively studied by Raman spectroscopy, which is one of the most powerful methods to prove both the vibration and electric band structure [9, 10, 11]. Kim et al., reported that the twisted (or rotated) BLG (tBLG), stacked rotationally by larger than 14 °, shows the SLG-like Raman spectra [8]. In addition, such phenomena have observed in rotationally faulted (rf-) multi-layer graphene (MLG) [12, 13, 14]. In order to facilitate these excellent properties, it is important to fabricate the thickness-controlled rf-MLG in high-quality and to establish the reliable characterization methods. In this paper, we present the growth of the thickness-controlled rf-MLG on the Ni foil by chemical vapor deposition (CVD) and its characterization by micro-Raman spectroscopy.

# 2. Experimental

rf-MLGs were grown on Ni foils by the low-pressure CVD. In order to control the thickness of rf-MLG, we used two different sets of the flow rates (50, 10, 100 sccm and 400, 10, 160 sccm for $CH_4$, $H_2$, Ar) and two different thicknesses of Ni foil (10, 50 μm) as in table 1. And the sample number is denoted by the growth conditions. The mixed gas was introduced into the CVD quartz chamber for 7 minutes in the vacuum of ~$10^{-1}$ Torr, following the annealing of the Ni substrate at 1000 °C for 1 hour in the vacuum of ~$10^{-6}$ Torr without gas flow. The sample was cooled down to room temperature by full-open of the furnace, retaining Ar flowing. Surface morphology of rf-MLG was investigated by scanning electron microscopy (SEM) images acquired by JSM-7000F (JEOL Ltd.). Thickness of rf-MLG was estimated by the combination of X-ray diffraction (XRD) and transmittance measurements by using Miniflex600 (Rigaku Ltd.) and home-built optical instruments equipped with CS260 monochromator (Oriel Ltd.), respectively. For transmittance measurements, rf-MLGs were transferred onto the glass substrate by the conventional process with assistance of the polymethyl methacrylate (PMMA). The rf-MLGs, grown on the Ni foil, were characterized by confocal micro-Raman spectrometer (NRS-5100, Japan optics Ltd.) equipped with the 532 nm laser whose spot size is under 1 μm and the spectrograph with the focal length of 300 mm. We used the grating of 600 lines/mm to obtain Raman spectra in the wide range, simultaneously.

## 3. Results and Discussion

Figure 1(a) and 1(b) are the SEM images of rf-MLGs (No.1−No.4) at 800× and 10000× magnifications, respectively. We can recognize that grains and white wrinkles are of different sizes in each sample. In the left images of rf-MLGs (No.1, No.3) grown on the Ni foil of 10 μm ($Ni_{10μm}$), the grains are as large as several tens μm and distinct, while the wrinkles are very thin and cannot be seen well at 800× magnification (Fig. 1(a)). The right images of rf-MLGs (No.2, No.4) on the Ni foil of 50 μm ($Ni_{50μm}$) show that grains are several times larger than those of rf-MLG on $Ni_{10μm}$, and the wrinkles are relatively thicker and are clearly observed over the grains. On the other hand, the upper (No.1, No.2) and the lower (No.3, No.4) images correspond to the surface of the rf-MLGs fabricated with the $CH_4$ of 50 sccm and 400 sccm, respectively (see the table 1). The wrinkle is clearer in No.4 than in No.2 in spite of the similar morphology. In Fig. 1(b), we can also observe the wrinkles on the surface of rf-MLG on $Ni_{10μm}$ (No.1, No.3) at 10000× magnification. In addition, there are white spots in No.1 and No.3. Such spots have often been observed in a graphene grown by CVD and are usually considered to be impurities from the substrate of metal catalyst. This implies that the sample surfaces of rf-MLG on $Ni_{10μm}$ are thinner than those of rf-MLG on $Ni_{50μm}$. The area surrounded by the wrinkle is larger and flatter in No.2 and No.4 than in No.1 and No.3. From the above results, the surface becomes flatter and the wrinkle is clearer as rf-MLG is getting thicker, that is, the layer number increases.

In order to investigate the crystallinity and the thickness of rf-MLG, the XRD and optical transmittance measurements were carried out. Figure 2(a) shows the XRD patterns of rf-MLGs (No.1−No.4). We observe a dramatic change of the peak intensity caused by the thickness of rf-MLG. There are typical graphite (002) peak at ~26.4 ° and Ni (200) peak at ~51.7 ° with strong intensity, and graphite (004) peak at ~54.5 ° and Ni (111) peak at ~44.4 ° with weak intensity. The existence of the graphite peaks indicates that the rf-MLG grown on Ni foil have been well stacked along the perpendicular direction to the surface plane. In No.1 and No.3, the intensity of graphite (002) peak is very weak compared to that of the Ni (200) peak. On the other hand, the graphite (002) peak is much larger than the Ni (200) peak in No.2 and No.4. These indicate that the thickness of rf-MLG on $Ni_{10μm}$ is much thinner than that of rf-MLG on $Ni_{50μm}$. Figure 2(b) shows the optical transmittance spectra of rf-MLGs (No.1−No.4) at normal incidence. All spectra increase with the wavelength of the incident light, showing the typical transmittance of MLG [5, 16]. Due to the stacking of a large number of graphene layers, transmittance is very low, compared to that of SLG. At 550 nm, the transmittances are ~13.6 % (No.1), ~2.2 % (No.2), ~13.5 % (No.3), and ~1.2 % (No.4). It is well known that SLG exhibits the optical absorption of ~2.3 % which is determined by the fine-structure constant due to interactions between the Dirac electron and the photon [16]. Moreover, the absorption of a few-layer graphene (FLG) is proportional to the number of graphene layers due to the weak bonding (van der Waals interaction) between the graphene layers. The optical transmittance of MLG has demonstrated by considering the interlayer hopping and different stacking order. It has been reported that the optical transmittance is not dependent on the stacking orders (AB- and ABC-stacking) and show good agreement between the numerical simulation and the experimental data [15]. Therefore, we employed the above method to estimate the thickness of rf-MLG by considering the independent stacking order and the weak van der Waals interaction of rf-MLG. The inset is the plot of the relation between the number of layers and the transmittance at 550 nm, using the equation $T = (1 + 1.13\pi\alpha N/2)^{-2}$ ($N$ is the number of graphene layer, and 1.13 is a correction coefficient for the deviation from the universal optical conductance ($e^2/4\hbar$)) [15]. The number of layers of No.1 and No.4 are estimated to be 132 and 634, respectively, which correspond to the thickness of ~44 nm and ~212 nm, respectively. From the above results of the SEM, XRD and transmittance measurements, we can conclude that the thickness of rf-MLG can be controlled effectively by the thickness of the Ni foil catalyst rather than the flow rate of $CH_4$, $H_2$, Ar.

Next, let us characterize the Raman spectra of the fabricated rf-MLGs. Figure 3(a) shows the representative Raman spectrum of rf-MLG on Ni foil (No.1, Pattern I in Fig. 4(a)). We observe two typical G and 2D Raman bands at ~1580 and ~2700 cm$^{-1}$, respectively. The G band has been assigned to the degeneracy of the in-plane transverse optic (TO) and the longitudinal optic (LO) phonon modes around the Γ-point ($E_{2g}$ symmetry). The 2D band has been considered to be associated with the second-order Raman scattering involving the TO phonon mode around the K-point (breathing vibration). It has been well known that the 2D band is largely enhanced by the double-resonance process in SLG compared to other graphitic materials. Therefore, the intensity ratio of the G and 2D bands ($I_{2D}/I_G$) has facilitated the estimation of the number of graphene layers. In general, $I_{2D}/I_G$ larger than 2 has been considered as an indication of SLG. In Fig. 3(a), the observed Raman spectrum reveals $I_{2D}/I_G$ ~ 3.7, showing the SLG-like behavior. However, the fabricated rf-MLG has the thickness of ~44 nm and the number of layers is about 132 as shown in Fig. 2. Such behaviors have been reported in rf-MLG, and a possible mechanism has been suggested: the rotational stacking of the graphene layers causes the weakness of the van der Waals bonding between the graphene layers, and then the 2-dimensional properties become stronger, showing the SLG-like behavior [25, 26]. Beyond the above two main peaks, we observe the small D band around 1350 cm$^{-1}$, which originates from defects. The intensity ratio of the D and G bands ($I_D/I_G$) is ~0.04, indicating that the fabricated rf-MLG is of high quality. In the enlarged plot (Fig. 3(b)), we can observe most of the overtone and combination modes allowed in sp$^2$-carbon materials as listed in table 2 [9, 10]. Here, we will confine our interest to the specific peaks to discuss the obtained Raman spectra. The $P_4$ (1500 cm$^{-1}$) and $P_6$ (1680 cm$^{-1}$) peaks are located around the G band. From the previous reports, $P_4$ consists of the R (TO) mode around 1430 cm$^{-1}$ [7] and the LO-ZO′ mode around 1520 cm$^{-1}$ [17, 18]. And, the $P_6$ peak includes the three modes; the D′ peak (LO) around 1620 cm$^{-1}$, the R′ (LO) peak around 1625 cm$^{-1}$, the LO+ZO′ mode around 1720 cm$^{-1}$ [17, 18]. Recently, in twisted graphene heterostructures, the R and R′ peaks were identified as $T_e$ (TO mode of inter-layer Raman scattering) and $L_a$ (LO mode of intra-layer Raman scattering), respectively [27]. Contribution of the D′ peak to the $P_6$ peak is very small because of the high-quality of the fabricated rf-MLG (very small intensity of D peak). In particular, the R and R′ peaks are related to the rotational stacking of the graphene layers in tBLG [7, 12, 13, 28]. The observed R and R′ peaks in rf-MLG are not clear comparing to those of tBLG, which may come from a large number of stacking with the different angles. We will discuss more about the R peak in Fig. 4 and Fig. 5. The $P_7$ peak at ~1750 cm$^{-1}$ is the overtone of ZO mode (2ZO) and usually assigned by the M band, which have information about the interaction between the graphene layers along the direction of the out-of-plane [20]. Here, it should be also noted that the M band is the infrared (IR) active and not shown in both SLG and tBLG [21]. In rf-MLG, the observation of the IR active peak indicates that the stacking along the out-of-plane is different from the usual AB- and ABC-stacking. Cong et al. have reported the M band can be Raman active by loosing the selection rule due to the rotational stacking [21]. The $P_8$ (~1880 cm$^{-1}$) and $P_9$ (~2035 cm$^{-1}$) peaks are combination of the in-plane modes, and are assigned to LO+TA and LO+LA, respectively [19, 21, 22, 23]. These peaks have been considered to come from the double resonance Raman process and used as indications of the two-dimensionality of the graphene.

Figure 4 shows the Raman spectra of No.1 in the range of 1170–2270 cm$^{-1}$, which are normalized to $I_G$ after the background subtraction of a constant value. They can be categorized by the four patterns as follows: (Pattern I) $I_{2D}/I_G$ is greater than two; (Pattern II) $I_{2D}/I_G$ is slightly greater than one and less than two; (Pattern III) $I_{2D}/I_G$ is slightly less than one; (Pattern IV) $I_{2D}/I_G$ is much less than one. In the inset, the Raman spectra of the 2D band were plotted using the y-axis of $I_{2D}/I_G$. As the pattern changes from I to IV, $I_{2D}/I_G$ decreases accompanying the variation of peak structure. In the pattern I, the 2D band is a single intense peak at ~2700 cm$^{-1}$. On the other hand, the 2D band exhibits the multi-peak structure with two main intensities at ~2700 cm$^{-1}$ and ~2720 cm$^{-1}$ in the patterns II–IV. As the pattern changes from II to IV, the spectral weight shifts from the low frequency (~2700 cm$^{-1}$) to the high frequency mode (~2720 cm$^{-1}$). These behaviors are very similar to the variation of the Raman spectra as the number of the graphene layer increases from monolayer to multilayer in the Bernal (AB-) stacking [6]. Moreover, we clearly observe the systematic variation of the out-of-plane ($P_7$) and in-plane ($P_8$, $P_9$) modes. With increasing $I_{2D}/I_G$, the in-plane peaks increase but the out-of-plane peak decreases oppositely. This is consistent with the previous reports on the tBLG and rf-MLG where the comparison of $I_{2D}$ and $I_G$ has successfully accounted for the dimensionality of the electronic properties [13, 21, 23]. In addition, we can recognize that the $P_2$ (~1230 cm$^{-1}$) and $P_{10}$ (~2220 cm$^{-1}$) peaks are also getting intensive with increasing $I_{2D}/I_G$, showing the similar behavior with the in-plane peaks ($P_8$, $P_9$). This means that the $P_2$ and $P_{10}$ peaks can be assigned to the in-plane modes. Actually, $P_{10}$ has been considered to be the combination mode of TO+TA in tBLG [22, 23]. It should be also noted that the observed behavior of the $P_{10}$ peak is different from the few-layer graphene of the Bernal (AB-) stacking, where the $P_{10}$ peak little varies [23]. When we consider the phonon dispersion of graphene [29, 30], the most possible mode, which corresponds to the peak frequency of $P_2$ (~1230 cm$^{-1}$), is the TO mode near K-point. Therefore, we tentatively assign $P_2$ to TO. Further studies are needed to clarify its Raman process.

Here, we should discuss about the relation between $I_{2D}/I_G$ and the rotational angle, too. It is well known that $I_{2D}/I_G$ depends on the rotational angle between the inter-layers of the graphene in tBLG: $I_{2D}/I_G$ is close to that of SLG as the rotational angle is larger than ~14 ° [8]. However, in rf-MLG, it is difficult to determine the rotational angle from $I_{2D}/I_G$, because of the large number of layers. From the similarity behaviors of the in- and out-of-planes between rf-MLG and tBLG, we can infer that the graphene layers are stacked by the weak van der Waals interaction, and the rotations of the top layers play an important role to determine the variation of $I_{2D}/I_G$.

Figure 5 shows the Raman spectra of No.1–No.4 by pattern, i.e. (a)(b) pattern I, (c)(d) pattern II, (e)(f) pattern III, and (g)(h) pattern IV. The left and right sides of Fig. 5 are the Raman spectra in the normal and enlarged scales, respectively. The Raman spectra of pattern I for all types of the sample (No.1–No.4) are displayed in Fig. 5(a). Their $I_{2D}/I_G$s are about 2.3–9.8, showing the SLG-like behavior. The full width at half maximum (FWHM) of the 2D band is from ~26 to ~38 cm$^{-1}$, which are comparable to those of SLG. In this scale, $I_D$ is very small and is not clearly seen in all samples, indicating little defects. Figure 5(b) shows the enlarged Raman spectra of the pattern I in the range of 1170–2270 cm$^{-1}$. It is found that the D bands of No.1–No.4 at ~1350 cm$^{-1}$ are of the similar intensity, indicating the similar sample quality. There is the small $P_2$ at ~1230 cm$^{-1}$, which is slightly larger in the No.2 and No.3 samples than in No.1 and No.4. This seems to be correlated with the fact that the $I_{2D}/I_G$ values are also larger in No.2 and No.3 than in No.1 and No.4. As mentioned in Fig. 4, the R peak near the G band reveals the rotational stacking of the adjacent graphene layers to each other. The R peak at ~1453 cm$^{-1}$ in No.4 is especially sharper comparing with that of the other samples. The R peak of No.2 is also clear but is located at the higher frequency (~1470 cm$^{-1}$) than that of No.4. It has been reported that the R peak is shifted to the lower frequency as the rotation angle increases [7, 28]. Therefore, it is expected that graphene layers near the top in No.4 are stacked by larger angle than those in No.2. In No.1 and No.3, the R modes appear as the shoulder at ~1470 cm$^{-1}$. Moreover, the small out-of-plane mode ($P_7$) and the large in-plane modes ($P_8$, $P_9$) indicate strong interactions in the in-plane of rf-MLG, i.e. the good two-dimensionality.

Figure 5(c) and 5(d) show the Raman spectra of the pattern II. It is found that the average FWHM of the 2D band is ~47 cm$^{-1}$ and is a little larger than that in the pattern I, indicating the increase of the band width of the Dirac-cone. Comparing with the pattern I, the prominent different point is the increase of the out-of-plane mode and the decrease of the in-plane modes, indicating the increase of the hybridization between the graphene layers. In addition, in Fig. 5(e), 5(f), 5(h) and 5(i), we find that, in the pattern III and IV, the in-plane peaks decrease and the out-of-plane increases, more and more. From the above results, we can safely conclude that the comparison of the in-plane and out-of-plane modes still plays an important role to estimate the two-dimensional properties in rf-MLGs of the different thickness. This means that the thickness does not affect the two-dimensional properties in rf-MLG. The D band also shows systematic variations. As the pattern changes from I to II and III, the intensity of the D band decreases and, eventually disappears in the pattern IV. This implies that the rotational stacking of graphene layers causes deformations of the honeycomb crystal structure of graphene.

Figure 6 shows the variation of the in-plane ($P_2$, $P_8$, $P_9$, $P_{10}$), out-of-plane ($P_7$) modes, and the 2D band ($P_{13}$) in the pattern I for No.1–No.4. In Fig. 6(a), the left y-axis is the peak intensity of the in-plane and out-of-plane modes ($I_{IN,OUT}$), and the right y-axis is that of the 2D band. All peaks were normalized to the intensity of the G band. As discussed in Fig. 4 and Fig. 5, the four in-plane modes show a similar tendency with the 2D band, while the out-of-plane mode shows the opposite behavior to the in-plane and 2D peaks. In Fig. 6(b), the intensities of the peaks (y-axis) are plotted as a function of the intensity of the 2D band (x-axis). All peaks are proportional to the 2D band by the different slopes. In order to investigate the relation between the above peaks and the 2D band, we normalized all peaks to the own largest value. This clearly reveals that the intensities of the in-plane modes are scaled to that of the 2D band, indicating they share the common mechanism, i.e. the double resonance Raman process [9, 19].

Finally, it is valuable to mention the distribution of each pattern in each sample. We measured the Raman spectra over the sample as evenly as possible. While the over half of the obtained spectra are pattern I or II in No.1, the less than 25 % of the spectra are pattern I or II in No.4.

## 4. Conclusions

We have fabricated rf-MLGs on Ni foils in the different thickness by the CVD method, and characterized them in detail by the Raman spectroscopy measurements. The surface morphology and thickness were investigated by the SEM, XRD, and transmittance measurements. The obtained data showed that the thickness of rf-MLG can be effectively controlled by the thickness of the Ni foil. In the Raman spectroscopy measurements, we successfully accounted for the dimensionality of rf-MLG by the analysis of the in-plane ($P_8$, ~1880 cm$^{-1}$; $P_9$, ~2035 cm$^{-1}$) and out-of-plane ($P_7$, ~1750 cm$^{-1}$) modes as in tBLG, in spite of many graphene layers and the different thickness. Here, we also attribute $P_2$ (~1230 cm$^{-1}$) and $P_{10}$ (~2220 cm$^{-1}$) to the in-plane modes due to the similar behavior of the typical in-plane modes ($P_8$, $P_9$). In addition, it is found that the peak intensities of the above four in-plane modes ($P_2$, $P_8$, $P_9$, $P_{10}$) are proportional to that of 2D band. This indicates that they share the common Raman resonance process. We believe that our growth method of the thickness controlled rf-MLG and the analysis of Raman spectra would provide useful information for both the understanding of fundamental properties and applications.

Table 1: Sample number and corresponding CVD parameters

| No. | Gas mixture (sccm) $CH_4, H_2, Ar$ | Thickness (μm) Ni-foil |
|---|---|---|
| 1 | 50, 10, 100 | 10 |
| 2 | 50, 10, 100 | 50 |
| 3 | 400, 10, 160 | 10 |
| 4 | 400, 10, 160 | 50 |

Table 2: Raman peaks in rf-MLG

| Peak No. | Raman shift (cm$^{-1}$) | Nomen-clature | Phonon modes | Note | Reference |
|---|---|---|---|---|---|
| P$_1$ | 1100 | D″ | LA | - | [11] |
| P$_2$ | 1230 | - | (TO) | (in-plane) | - |
| P$_3$ | 1350 | D | TO | defect | [9, 10] |
| P$_4$ | 1500 | R, - | TO, LO-ZO′ | - | [7, 17, 18, 19] |
| P$_5$ | 1580 | G | TO and LO | - | [9, 10] |
| P$_6$ | 1680 | R′, D′, - | LO, LO, LO+ZO′ | - | [17, 18, 19] |
| P$_7$ | 1750 | M | 2ZO | out-of-plane | [20, 21] |
| P$_8$ | 1880 | - | LO+TA | in-plane | [21, 22, 23] |
| P$_9$ | 2035 | - | LO+LA | in-plane | [21, 22, 23] |
| P$_{10}$ | 2220 | - | TO+TA | in-plane | [22, 23] |
| P$_{11}$ | 2330 | - | - | N$_2$ gas | - |
| P$_{12}$ | 2453 | D+D″ | TO+LA | - | [9, 10] |
| P$_{13}$ | 2700 | 2D | 2TO | - | [9, 10] |
| P$_{14}$ | 2900 | G+D | TO | - | [9] |
| P$_{15}$ | 2970 | D+D′ | TO+LO | - | [9] |
| P$_{16}$ | 3175 | 2G | 2TO and 2LO | - | [9] |
| P$_{17}$ | 3245 | 2D′ | 2LO | - | [9] |

* ( ) indicates the results obtained in this work.

* ZO′ is the layer-breathing mode.

* The TO+LA mode, expected at ~1980 cm$^{-1}$ [22], could not be clearly observed due to small intensity.

* P$_{11}$ is the vibration mode of N$_2$ gas on rf-MLG [24].

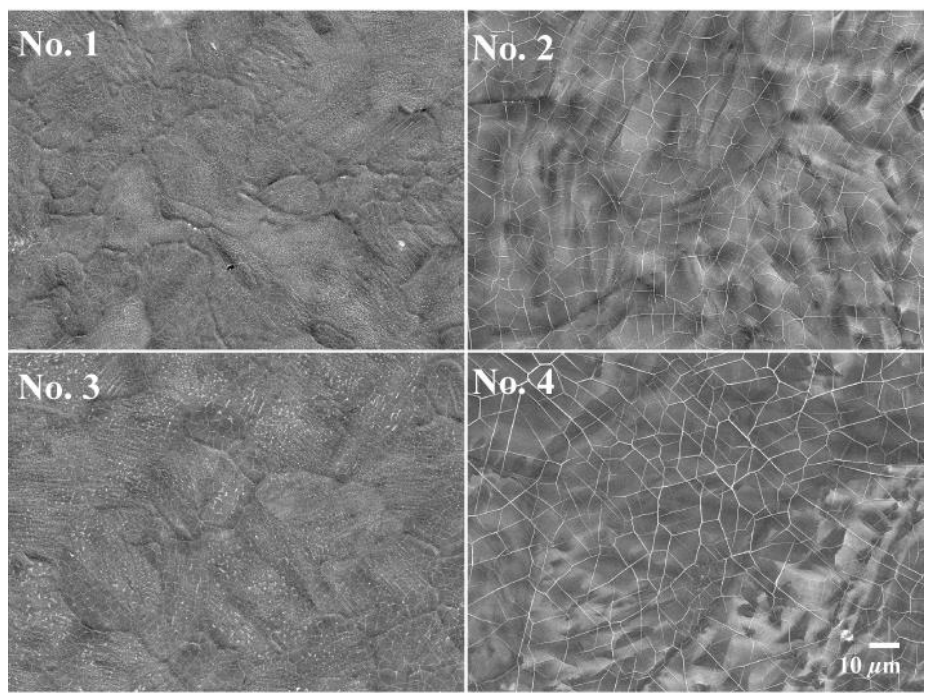

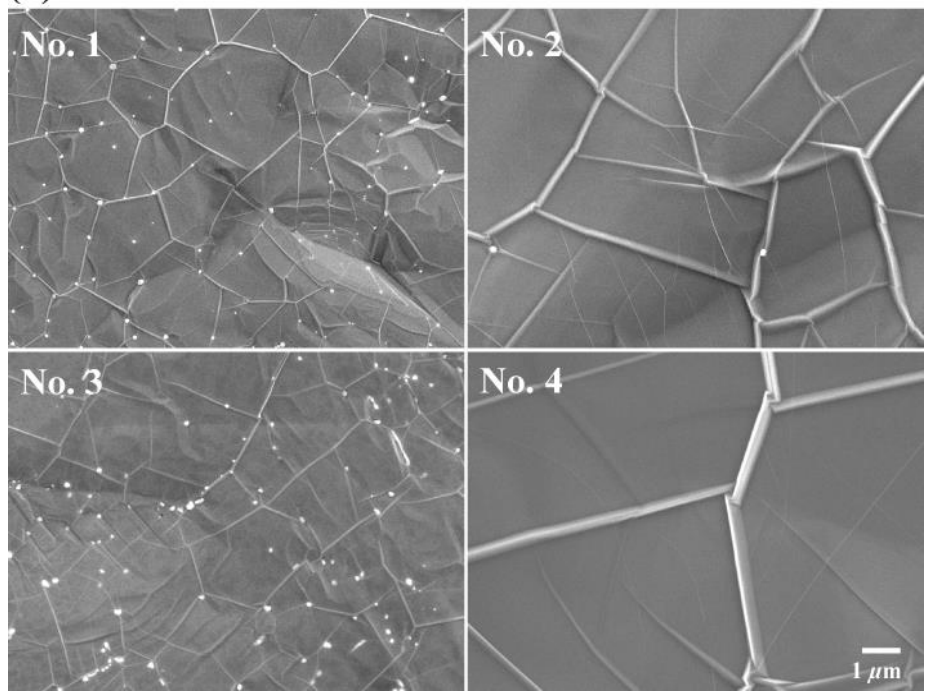

Figure 1: SEM images of rf-MLGs (No.1−No.4) at (a) 800× and (b) 10000× magnifications.

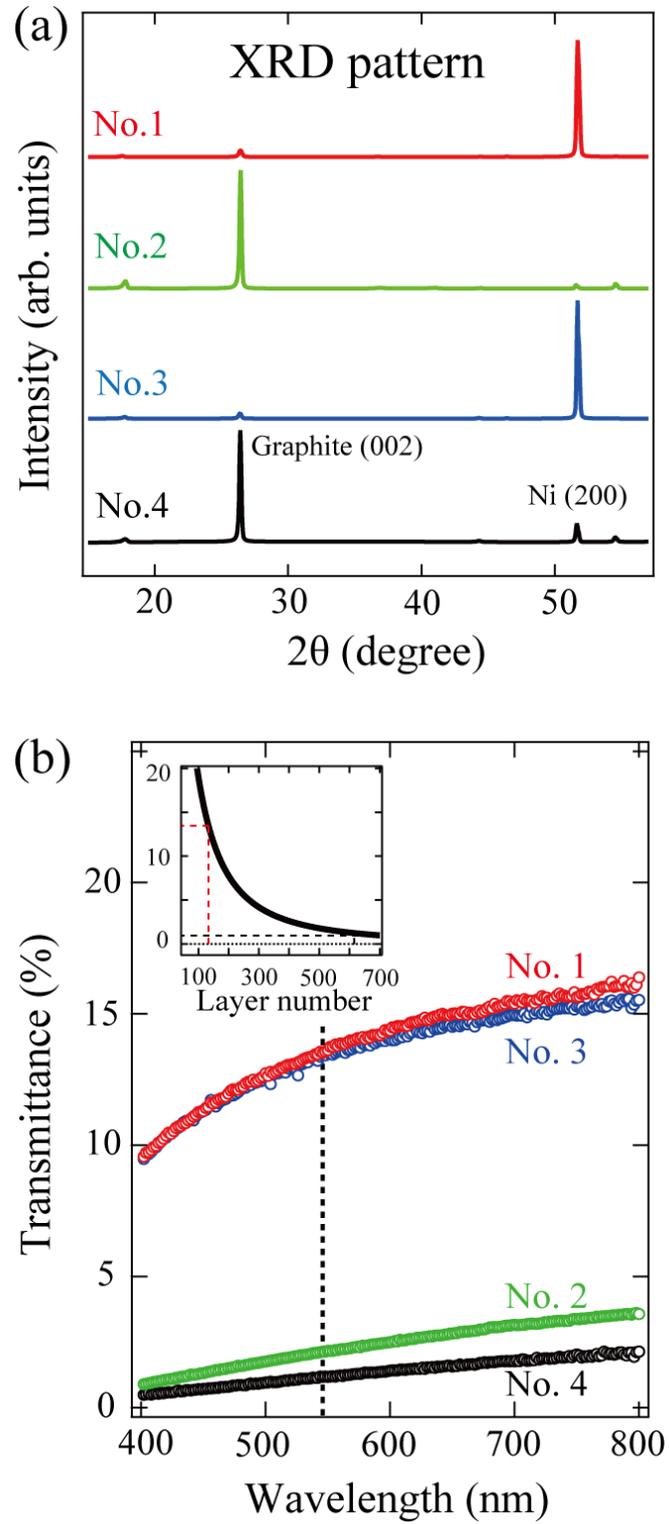

Figure 2: (a) XRD patterns and (b) Transmittance spectra of rf-MLGs (No.1–No.4). The inset is a plot of relation between the layer number and the transmittance at 550 nm (No.1 and No.4) [15]. Dashed-lines are guides to the eye.

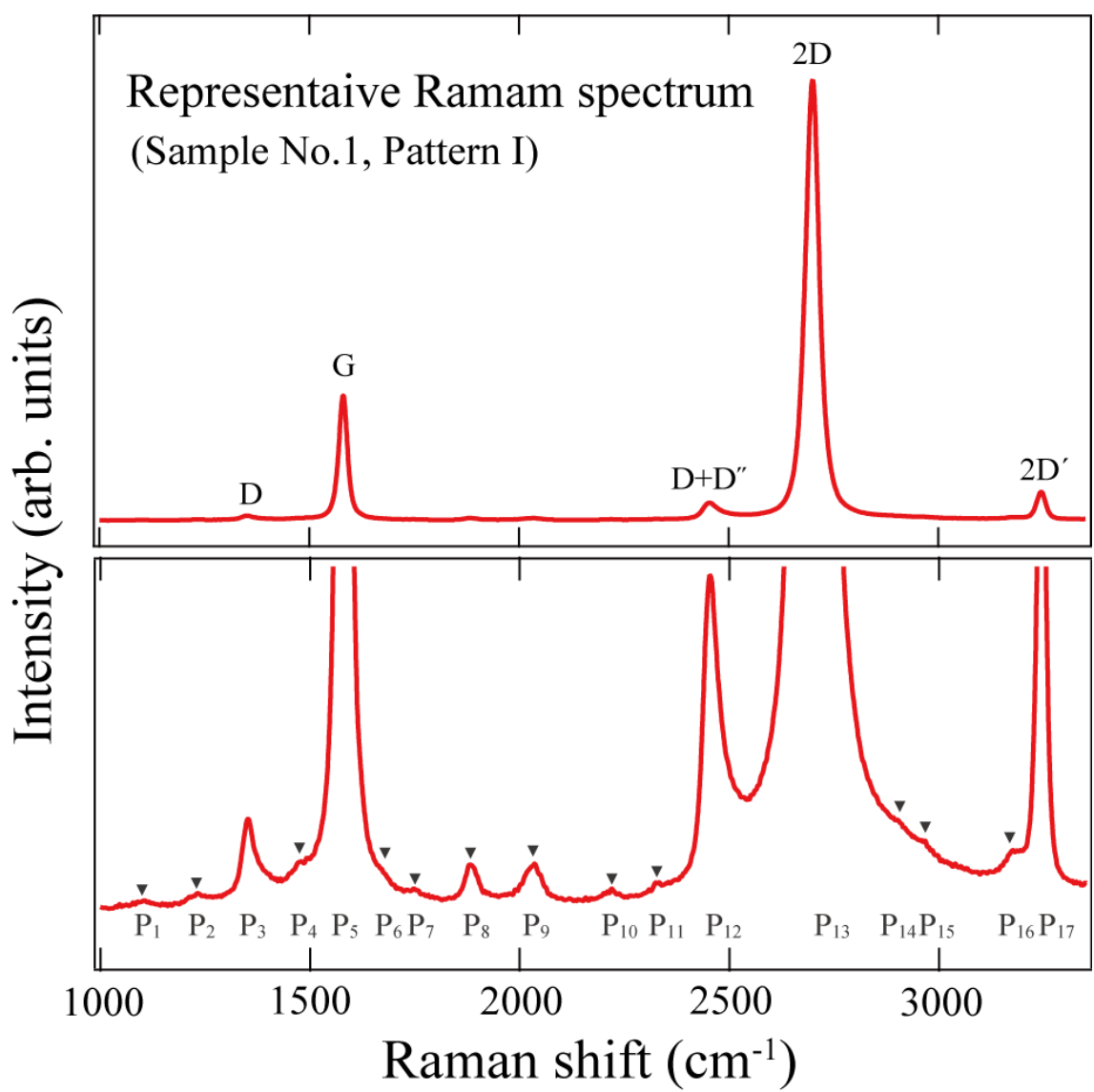

Figure 3: (a) Representative Raman spectrum of rf-MLG (No.1, Pattern I). (b) Enlarged Raman spectrum in the low intensity regime.

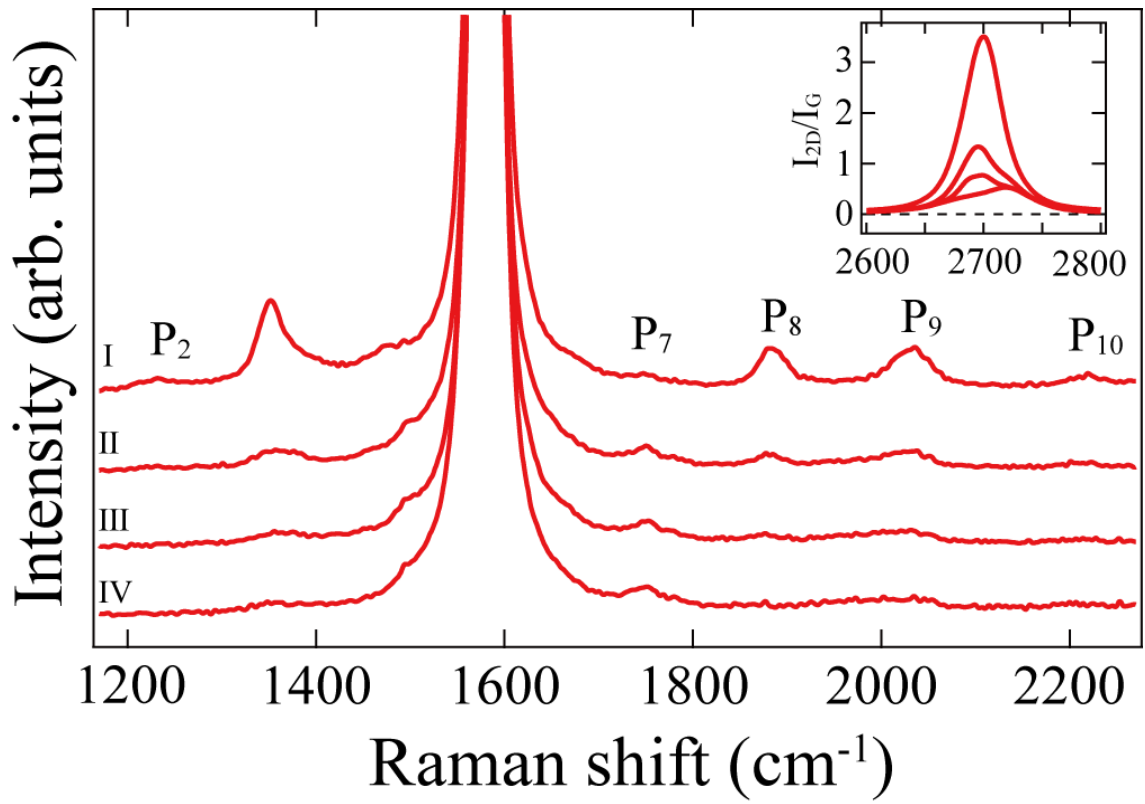

Figure 4: Raman spectra of the four patterns (I–IV) in rf-MLG (No.1).

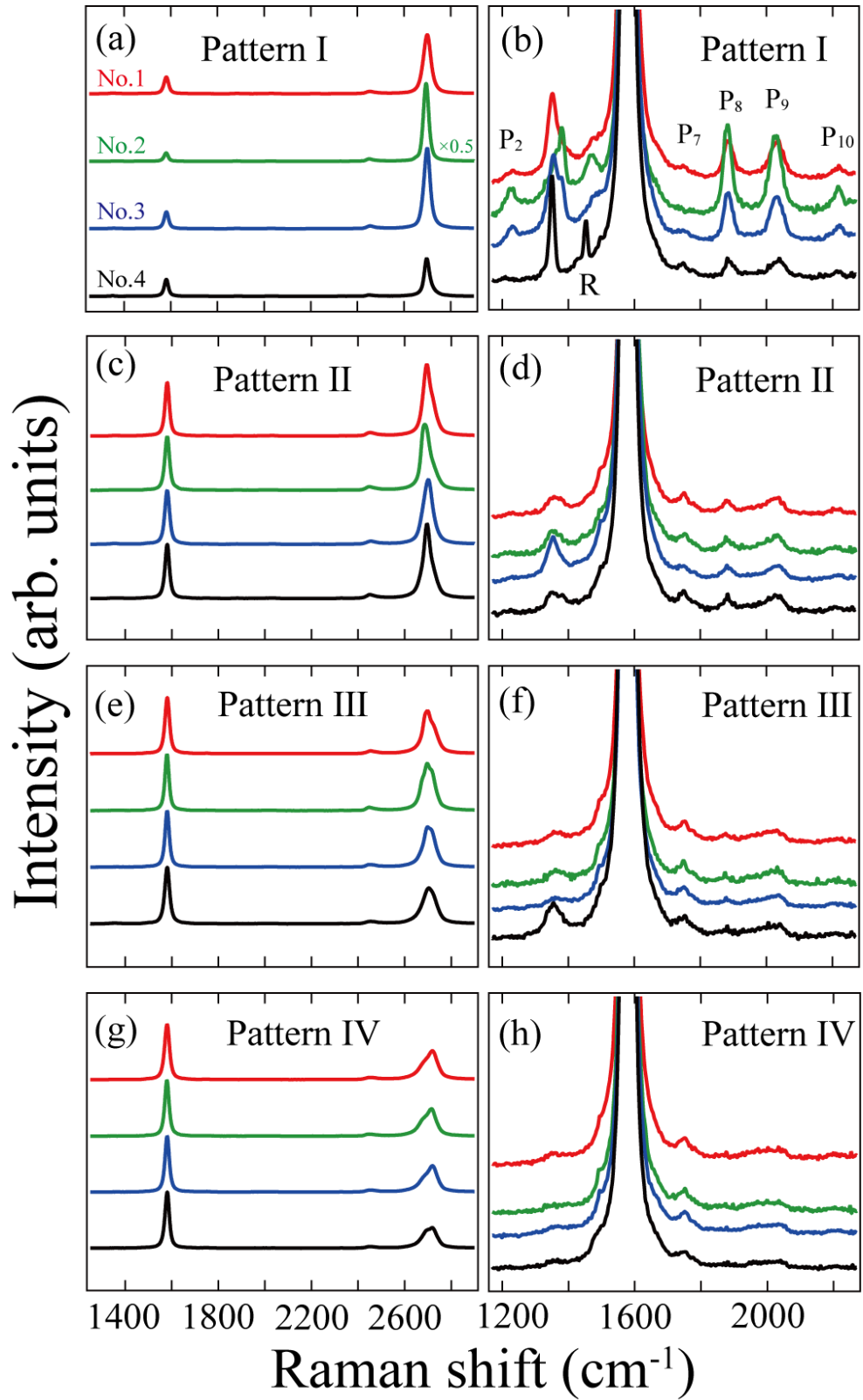

Figure 5: Raman spectra of pattern I (a,b), II (c,d), III (e,f), and IV (g,h) for each rf-MLG (No.1–4).

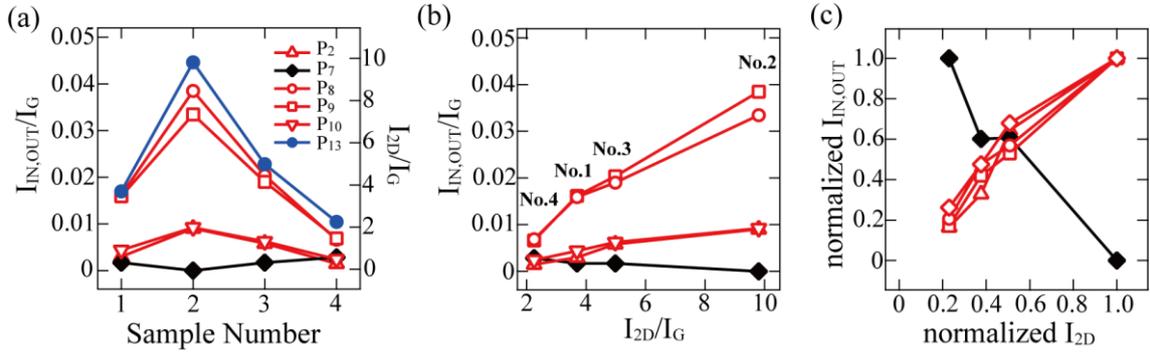

Figure 6: Comparison of the peak intensities of the in-plane ($P_2$, $P_8$, $P_9$, $P_{10}$), out-of-plane ($P_7$) modes, and the 2D band ($P_{13}$) (Pattern I for No.1–No.4).